\documentclass[twocolumn,showpacs,preprintnumbers,amsmath,amssymb] {revtex4}
\usepackage{graphicx}
\def\be{\begin{equation}}
\def\ee{\end{equation}}
\def\bee{\begin{eqnarray}}
\def\eee{\end{eqnarray}}

\begin{document}
\title{On the radial propagation of turbulence in gyro-kinetic toroidal systems}

\author{P. Migliano$^1$, R. Buchholz$^1$, S.R. Grosshauser$^1$, W.A. Hornsby$^1$, A.G. Peeters$^1$}

\vskip 0.2 truecm 

\address{$^1$ University of Bayreuth, Physics department, Universit\" atsstra{\ss}e 30 Bayreuth, Germany}


\begin{abstract}

In this paper a conservation equation is derived for the radially dependent entropy in toroidal geometry
using the local approximation of the gyro-kinetic framework.
This equation naturally leads to an operative definition for the turbulence intensity. 
It is shown that the conservation equation can be split in two separate conservation equations, one 
describing the dynamics of the zonal modes and one for the non-zonal modes. 
In essence the paper provides an operative tool for both analytic as well as numeric studies of the 
radial propagation of turbulence in tokamak plasmas. 

\end{abstract}

\maketitle

\section{INTRODUCTION}\label{INTRODUCTION}

A detailed understanding of turbulent transport in magnetically confined plasmas is 
essential for the development of nuclear fusion devices.
One of the key questions regards the relation between local and global model descriptions
of plasma turbulence.   
A fundamental issue of the latter research area is understanding the role of the turbulent transport of 
turbulence intensity (turbulence spreading) that occurs in the global model, but is lacking in 
any local description. Several authors have considered this problem in the past. 
In \cite{GAR94} a fluid model is used to show that mode coupling provides an efficient mechanism 
for the radial propagation of turbulence in tokamaks.
Furthermore, a conservation equation, for the evolution of the local intensity $I$ of the turbulence, 
is given in \cite{HAH04} in the form of a Fisher-Kolmogorov equation \cite{FIS37, KOL37} with 
an inhomogeneous diffusion coefficient. In the case of weak turbulence (see \cite{HOR99}) it takes the form 
\be
{\partial I \over \partial t} - {\partial \over \partial \psi} \biggl [ D (I) {\partial I \over \partial \psi} \biggr ] = 
\gamma I -  k_\perp^2 I^{2} \ .
\label{HDLII} 
\ee
An argument on the dynamics of turbulence spreading, which gives validity to this equation, can be found in \cite{GUR05, GUR06}. 
In the following we give a brief description of the terms that appear in the equation. The second term on the left hand side describes the spatial scattering of turbulence energy induced by non-linear coupling. The local turbulent diffusion coefficient $D(I)=D_0I$ is considered to be proportional to the intensity itself. The local growth rate of the intensity is $\gamma$. 
The non-linear saturation of the turbulence is modelled by the non-linear damping term $-k_\perp^2I^2$, where $k_\perp$ is a suitably chosen scale of  
the turbulent fluctuations. 
The variable $\psi$ is the radial coordinate, and $t$ is the 
time. 

Eq.~\eqref{HDLII} has been largely used by these and other authors (see for example
\cite{WAL05} and \cite{WAN11}) to tackle the problem of turbulence spreading. 
It provides a very useful model for the discussion of turbulence spreading, 
but it is affected by some deficiencies. 
Although physically motivated, it is not derived from first principles. 
Indeed, the evolution of the local turbulence intensity defined as the squared modulus of the electrostatic 
potential can be shown not to satisfy a conservation equation of the form given. 
Therefore, numerical calculations can not be directly interpreted in terms of the dynamics described by this equation. 
Another point of concern is that there is no clear separation between turbulent and zonal intensity. 
With the latter we refer to the potential perturbation connected with the zonal ($n = 0$ toroidal) mode. 
It is not obvious if the turbulent intensity should contain (or not contain) the zonal contribution. 
The points raised above provide a motivation to investigate the possibility of deriving analytically a conservation 
equation of the form given by Eq.~\eqref{HDLII}.  
In this paper we undertake this task starting from the gyro-kinetic framework. 
The goal is to give a solid foundation to the discussion of turbulence spreading, and to derive analytic 
expressions for the form of the turbulent flux of turbulence intensity. 

Our starting point is the choice of a quantity describing the intensity of the turbulence. 
A reasonable candidate is the entropy of the system, since the entropy is a measure of the departure from equilibrium, 
and the entropy satisfies a proper conservation equation.  
The idea of using the entropy to define the intensity of the turbulence has been already suggested in the 
literature, for instance in \cite{GUR13} where a balance equation for the entropy density is given starting 
from the drift kinetic equation in cylindrical geometry. 
Total (i.e. integrated over the entire computational domain) entropy conservation in the gyrokinetic toroidal system 
has also been extensively studied (see for instance \cite{CAN06}) for the case of the local limit approximation 
\cite{BEE95}. 
In this paper we perform a calculation close to the one given in \cite{CAN06}, in contrast 
we exclude the integral over the radial coordinate in order to explicitly keep track of the radial dependence of 
the perturbations. This procedure leads to an equation for the evolution of the radial dependent entropy of the 
system considered. 
The form of the conservation equation for the entropy leads naturally to an operative definition for the intensity 
of the turbulence and its conservation equation. 
Further analysis allow this equation to be split in two separate equations, one describing the dynamics of the 
turbulence intensity in the zonal ($n=0$ toroidal) mode and the other the turbulence intensity in the perturbations (non-zonal $n\ne0$ toroidal modes). 
The symmetry and simplicity of the resulting system of equations give a genuine insight into the connection between 
the dynamics of zonal and non-zonal modes.

\section{TURBULENCE INTENSITY BALANCE IN GYRO-KINETIC THEORY}\label{TURBULENCE_INTENSITY_BALANCE_IN_GYROKINETIC_THEORY}

In this section we derive the conservation equation for the radially dependent turbulence intensity of a collisionless plasma with no rotation, in the electrostatic case, 
for general toroidal geometry. 
The calculation is performed in the local limit approximation \cite{BEE95}, in particular we consider the case in which background quantities do not vary across the perpendicular (to the magnetic field) extent of the domain, applying periodic boundary conditions on the entire domain, excluding the integral over the radial coordinate. 
This choice, although it does not describe the most general case, allows to study the behaviour of radial inhomogeneities in the perturbations of the system.   

We need an operative definition for the intensity of the turbulence, i.e.
we look for a quantity which is radially dependent and satisfies a conservation equation in the form of Eq.~\eqref{HDLII}, so that numerical results from gyrokinetic 
simulations can then be properly interpreted in terms of the dynamics described by this equation.
As already pointed out in the introduction, a natural candidate is the entropy of the system. 
We define the radially dependent entropy of the particles of the $sp$-species as
\be
\epsilon_{sp} = -\int dx \ dv \ f_{sp}^{tot}\ln\frac{f_{sp}^{tot}}{F_M} \ ,
\label{entropy} 
\ee 
the radially dependent entropy of all particles is obviously obtained taking the sum over all species. 
In Eq.~\eqref{entropy} we have $dx \ dv=J \ ds \ d\zeta \ dv_\parallel \ d\mu$ with $J=\sqrt{g}$ Jacobian of the transformation ($g$ being the 
determinant of the metric tensor) and $f_{sp}^{tot}=F_M+ f_{sp}$ is the total distribution of the $sp$-species written as a sum of $F_M$, the 
equilibrium Maxwell distribution as given in Eq.~(66) of \cite{PEE09}, and a small perturbation $f_{sp}$ of order $\rho_*$ 
to the equilibrium (where $\rho_*=\rho/R$ is the normalized reference Larmor radius, with $R$ the tokamak reference major radius and $\rho = \sqrt{2 T / m} /\omega_{c}$ where $m$ is the reference mass, $\omega_{c}$ is the 
reference cyclotron frequency, $T$ is the reference temperature). 
We use gyrocenter field aligned Hamada coordinates ($\mathbf{X},v_\parallel,\mu$), $\mathbf{X}=(\psi,s,\zeta)$ being the 
gyrocenter position (with $\psi,s$ and $\zeta$ respectively radial, field line and binormal coordinates), $v_\parallel$ 
the parallel (to the magnetic field) velocity, and $\mu$ the magnetic moment $\mu=m_{sp}v_\perp^2/(2B)$ where $v_\perp$ is 
the velocity component perpendicular to the equilibrium magnetic field, $m_{sp}$ the mass of the $sp$-species and $B$ the background magnetic field 
strength.
We choose the Maxwell distribution $F_M$ as the reference distribution in the definition of the entropy to make the 
maximum entropy state correspond to the physical equilibrium distribution ($\epsilon_{sp}$ has a maximum when $f_{sp}=0$). 
It is important to stress again that in contrast to \cite{CAN06} here the integral is performed over the 
phase space excluding the radial coordinate $\psi$ in order to explicitly keep track of the radial dependence of the perturbations.  

We make a Taylor expansion of Eq.~\eqref{entropy} to the second order in $\rho_*$ then the following approximation holds 
\be
\epsilon_{sp} \approx -\int dx \ dv \left( f_{sp} + \frac{f_{sp}^2}{2F_M} \right) \ ,
\label{entropy_approx} 
\ee 
note that the first term does not vanish in this case since the integral is not performed over the entire phase space.
We build the equation which describes the time evolution of $\epsilon_{sp}$ using the gyrokinetic equation given in Eq.~(69) 
of Ref.~\cite{PEE09}, considering the case of a plasma as described at the beginning of this section. 
The time derivative of the first term in Eq.~\eqref{entropy_approx} simply gives 
\be
\int dx \ dv \left[\frac{\partial f_{sp}}{\partial t} + \frac{\partial}{\partial \psi}\left(f_{sp}v_E^\psi\right)\right] = 0 \ ,
\label{mass_density_conservation} 
\ee 
i.e. the continuity equation for the mass density, here $\psi$ is the radial coordinate, $t$ is the 
time and $v_E^\psi$ is the radial component of the perturbed $\mathbf{E}\times\mathbf{B}$ velocity. 
The time derivative of the second term of Eq.~\eqref{entropy_approx} can be rewritten in the form
\be
\int dx \ dv \ \frac{\partial}{\partial t}\left( \frac{f_{sp}^2}{2F_M} \right) = \int dx \ dv \left(\frac{f_{sp}}{F_M}\frac{\partial f_{sp}}{\partial t} \right) \ ,
\label{time_der_second_term} 
\ee
therefore we find
\be
\begin{split}
\int dx \ dv &\left[ \frac{\partial}{\partial t}\left( \frac{f_{sp}^2}{2F_M} \right) + \frac{Z_{sp}}{T_{sp}}\chi\frac{\partial f_{sp}}{\partial t}\right] = \\
= - & \int dx \ dv \ \frac{\partial}{\partial \psi}\left[\left( \frac{f_{sp}^2}{2F_M} + \frac{Z_{sp}}{T_{sp}}\chi f_{sp}\right)v_E^\psi\right] + \\
&+ \int dx \left[\left( \frac{1}{L_n}-\frac{3}{2}\frac{1}{L_T}\right)J_{sp} + \frac{1}{L_T}K_{sp}\right] \ ,
\end{split}
\label{intensity_conservation1} 
\ee 
where $Z_{sp}$ and $T_{sp}$ are respectively the electric charge and the  
temperature of the $sp$-species, $\chi=G(\phi)$ is the gyroaveraged perturbed electrostatic potential ($G$ is the gyroaverage operator and $\phi$ the perturbed electrostatic potential), $1/L_n$ and $1/L_T$ are the inverse density and temperature background gradient lengths, $J_{sp}$ and $K_{sp}$ are given by
\be
\begin{split}
J_{sp} &= \int dv \left(h_{sp} v_E^\psi\right) \\ 
K_{sp} &= \int dv \left(\frac{m_{sp}v^2}{2} \ h_{sp} v_E^\psi\right) \ ,
\end{split}
\label{particle_heat_flux_non_ad} 
\ee 
where $h_{sp}=f_{sp}+(Z_{sp}/T_{sp})\chi F_M$ is the $sp$-species non-adiabatic gyrocenter response, $m_{sp}$ the mass of the $sp$-species and $v^2=v_\parallel^2+v_\perp^2$ with $v_\parallel$ and $v_\perp$ velocity space coordinates as defined at the beginning of this section. 

Eq.~\eqref{intensity_conservation1} does not quite show the features of a proper conservation equation in the form of Eq.~\eqref{HDLII}, the problem is clearly the second term in the first line which requires particular attention. In the following we discuss how to deal with it. When integrating over the entire phase space a proper scalar product between functions of the gyrocenter coordinates can be defined, therefore the following relation holds exactly 
\be
\int d\psi \ dx \ dv \left[ G(s)t\right] =\int d\psi \ dx \ dv \left[ sG(t)\right] \ ,
\label{gyro_adj_entire_ps} 
\ee
where $s$ and $t$ are any functions of the gyrocenter coordinates and the hermiticity of the gyroaverage operator $G=G^{\dagger}$ (with $G^{\dagger}$ adjoint gyroaverage operator) has been used because of the local limit approximation (this identity is the analogous to Eq.~(28) in \cite{CAN06}). In our case Eq.~\eqref{gyro_adj_entire_ps} can not be directly applyed since the integration over the radial coordinate is not performed, but using periodic boundary conditions we can write
\be
\int dx \ dv \left[G(s)t\right] = \int dx \ dv \left[sG(t)\right] + \frac{\partial}{\partial \psi}\left(\Gamma_{GA}\right) \ ,
\label{gyro_adj} 
\ee
with $\Gamma_{GA}$ a periodic function of the radial coordinate only; its physical meaning will be soon clarified. 

We now manipulate the second term in the first line of of Eq.~\eqref{intensity_conservation1} according to Eq.~\eqref{gyro_adj}, then we use the quasineutrality condition written in the form
\be
\sum_{sp}\int dv \left[Z_{sp}G(f_{sp}) + \frac{Z_{sp}^2F_M}{T_{sp}}\left(G\left(\chi\right) - \phi\right)\right] = 0 \ ,
\label{quasineutrality} 
\ee 
and combining eqs. \eqref{mass_density_conservation} and \eqref{intensity_conservation1} we can write 
\be
\frac{\partial}{\partial t}(\epsilon + w) + \frac{\partial}{\partial \psi}\left(\Gamma + \Gamma_{GA}\right) + C = 0 \ ,
\label{entropy_balance} 
\ee 
where we have defined  
\be
\begin{split}
\epsilon &= \sum_{sp}\epsilon_{sp} \\
w &= \sum_{sp}w_{sp} = \sum_{sp}\int dx \ dv \ \frac{Z_{sp}^2F_M}{2T_{sp}^2}\left(\chi^2 - \phi^2\right) \\
\Gamma &= -\sum_{sp}\int dx \ dv \left[\left( f_{sp} + \frac{f_{sp}^2}{2F_M} + \frac{Z_{sp}}{T_{sp}}\chi f_{sp} \right)v_E^\psi\right] \\
C &= \sum_{sp}\int dx \left[\left( \frac{1}{L_n}-\frac{3}{2}\frac{1}{L_T}\right)J_{sp} + \frac{1}{L_T}K_{sp}\right] \ ,
\end{split}
\label{entropy_balance_terms_def} 
\ee 
while $\Gamma_{GA}$ is the term arising from leaving out the integration over $\psi$ when performing the operation in Eq.~\eqref{gyro_adj} with the gyroaverage operator. It is interesting to notice that because of Eq.~\eqref{quasineutrality} it is not possible to write a conservation equation for the entropy of one species ($\epsilon_{sp}+w_{sp}$), the conserved quantity is the entropy of the whole system.

Since Eq.~\eqref{entropy_balance} appears in the proper form of a conservation equation we can read out of it the physical meaning of each single term: $\epsilon + w$ is the radially dependent entropy of the system, with $\epsilon$ entropy in the particles and $w$ entropy in the electrostatic field; $\Gamma+\Gamma_{GA}$ is the radial flux of entropy, this means that the physical effect of Eq.~\eqref{gyro_adj} is giving rise to an additional contribution to the radial flux, the last term $C$ represents sources and sinks as fluxes in the background gradients. 

The contribution of $\Gamma_{GA}$ can be shown to be of higher order in the Larmor radius compared to $\Gamma$ as follows: by approximating the gyroaverage operator as 
\be
G \approx 1 -\frac{1}{4}\rho_*^2\Delta \ ,
\label{gyroaverage_operator_approx} 
\ee  
where $\Delta$ is the normalized Laplacian operator, and applying for each species the gyrokinetic ordering 
\be
\frac{f_{sp}}{F_M}\approx\frac{Z_{sp}\phi}{T_{sp}}\approx\rho_* \ ,
\label{gyrokinetic_ordering} 
\ee  
it is staightforward to show that 
\be
\Gamma_{GA} \approx \rho_*\Gamma \ .
\label{neglect_Gamma_GA} 
\ee 
We can therefore neglect the contribution of $\Gamma_{GA}$ to the total flux of entropy. Furthermore our purpose is to find a balance equation whose form can be directly related to Eq.~\eqref{HDLII} in the context of gyrokinetic theory, thus we quantitatively miss a small part of the radial flux but it does not qualitatively destroy the form of the balance equation.    

From now on, for simplicity in the notation, we omit the sum over the species and we get rid of the $sp$-index, however each quantity in the equations has to be understood as related to a particular species and the physical equations are obtained performing the sum over all species in the system. 

We consider Eq.~\eqref{entropy_balance} and subtract from it the continuity equation for the mass density \eqref{mass_density_conservation}, neglecting the contribution of $\Gamma_{GA}$ we are left with a conservation equation of the form
\be
\frac{\partial I}{\partial t} + \frac{\partial \Gamma_I}{\partial \psi} = C  \ ,
\label{intensity_new_conservation} 
\ee   
for the quantity
\be
I = \int dx \ dv \ \left[\frac{f^2}{2F_M}+\frac{Z^2F_M}{2T^2}\left(\phi^2 - \chi^2\right)\right] \ ,
\label{intensity_new} 
\ee   
where $\Gamma_I$ is given by
\be
\Gamma_I = \int dx \ dv \left[\left( \frac{f^2}{2F_M} + \frac{Z}{T}\chi f \right)v_E^\psi\right] \ ,
\label{flux_turbulence} 
\ee 
and $C$ given in Eq.~\eqref{entropy_balance_terms_def}. 
It is clear that Eq.~\eqref{intensity_new_conservation} has the same form as Eq.~\eqref{HDLII}, 
i.e. term by term starting from the left we have the time derivative of $I$, the radial derivative of 
its radial flux and the source terms. 
Furthermore, the quantity in Eq.~\eqref{intensity_new} is quadratic in the perturbation. 
For these reasons we choose $I$ as definition for the intensity of the turbulence. 

Although the intensity $I$ satisfies a conservation equation, it does still contain the zonal perturbation. 
Below we will, therefore, split the intensity in a zonal and a non-zonal contribution. 
The binormal coordinate $\zeta$ is an ignorable coordinate, therefore it can be treated spectrally. 
Using the Parseval's theorem the integral over $\zeta$  can be replaced by a sum over the toroidal modes ($n$), 
then the definition of the turbulence intensity given in Eq.~\eqref{intensity_new} is
\be
I = \int d\sigma \sum_n \left[\frac{|f_n|^2}{2F_M}+\frac{Z^2F_M}{2T^2}\left(|\phi_n|^2 - |\chi_n|^2\right)\right] \ ,
\label{intensity_zeta_spectral} 
\ee
where the sum runs over all integers $n \in (-\infty,+\infty)$, $d\sigma=J \ ds \ dv$ is a short-hand notation 
for the reduced infinitesimal volume of integration and all quantities with the subscript '$n$' are defined by 
their Fourier transform in the binormal direction, i.e. for a generic function $t$ of the coordinates we have 
\be
t(\psi,\zeta,s) = \sum_n t_n(\psi,s)e^{ik_n\zeta} \qquad k_n=\frac{2\pi n}{L_\zeta} \ ,
\label{fourier_repr_zeta}
\ee
with $L_\zeta$ length of the $\zeta$ domain in real space, then the quantity $|t_n|^2=t_nt_n^*$ is the square modulus of the complex Fourier amplitude $t_n$ with the star indicating the complex conjugate.

The expression in Eq.~\eqref{intensity_zeta_spectral} contains the contribution from the zonal modes $(n=0)$ and all other perturbations $(n\ne0)$. The zonal modes, including the contribution of the Geodesic-acoustic mode (GAM), might show a more complex 
behaviour (see for instance \cite{CHE04,MIK10}). Therefore we split the case of the zonal modes and all other perturbations writing the turbulence intensity in the form $I=I_{ZM}+I_P$ with
\be
\begin{split}
&I_{ZM} = \int d\sigma \left[\frac{|f_0|^2}{2F_M}+\frac{Z^2F_M}{2T^2}\left(|\phi_0|^2 - |\chi_0|^2\right)\right] \\
&I_{P} = \int d\sigma \sum_{n\ne0}\left[\frac{|f_n|^2}{2F_M}+\frac{Z^2F_M}{2T^2}\left(|\phi_n|^2 - |\chi_n|^2\right)\right] \ .
\end{split}
\label{intensity_split} 
\ee
The turbulence intensity flux $\Gamma_I$ can be written in the spectral representation for $\zeta$ using the Parseval's theorem together with the convolution theorem, we obtain
\be
\Gamma_I = \int d\sigma\sum_{n,m} \left[\left( \frac{f_m}{2F_M} + \frac{Z}{T}\chi_m \right)f_{n-m}\alpha_n^*\right] \ ,
\label{flux_turbulence_zeta_spectral} 
\ee 
where we have renamed the Fourier transform in the binormal direction of the $\mathbf{E}\times\mathbf{B}$ velocity as $\alpha_n=(v_E^\psi)_n$ in order to lighten the notation.
Unrolling now the sums over $n$ and $m$ in Eq.~\eqref{flux_turbulence_zeta_spectral} and using the identities 
\be
\alpha_{0}=0 \qquad \int d\sigma\sum_n\chi_n\alpha_n^*=0 \ ,
\label{identity_chi_alpha}
\ee
which hold due to the fact that $\alpha_n\propto ik_n\chi_n$, one can see that the flux $\Gamma_I$ can be split similarly to the turbulence intensity in the form $\Gamma_I = \Gamma_{I_{ZM}} + \Gamma_{I_{P}}$ where
\be
\begin{split}
&\Gamma_{I_{ZM}} = \int d\sigma \left[ \frac{h_0}{F_M} \sum_{n}\left(f_n\alpha_n^*\right)\right] \\
&\Gamma_{I_{P}} = \int d\sigma \sum_{\substack{n\ne m\\m\ne0}}\left[\left( \frac{f_m}{2F_M} + \frac{Z}{T}\chi_m \right)f_{n-m}\alpha_n^*\right] \ ,
\end{split}
\label{flux_turbulence_split} 
\ee
where $h_0=f_0+(Z/T)\chi_0F_M$ is the zonal modes component of the non-adiabatic gyrocenter response; the sum in the second term is performed over both indices $n$ and $m$ with the restrictions of $n\ne m$ and $m\ne0$. 

Eq.~\eqref{flux_turbulence_split} can be considered the main result of this work, in fact it is shown for the first time that the turbulence intensity flux can be split in two terms, one of which (the second one) does not contain any contribution from the zonal modes. 

The physical meaning of this property can be understood as follow: considering the relation
\be
\int dx \ dv \left(\frac{f_0}{F_M}\frac{\partial f}{\partial t} \right) = \int d\sigma \ \frac{\partial}{\partial t}\left( \frac{f_0^2}{2F_M} \right) \ ,
\label{time_der_second_term_zonal_modes} 
\ee
it is clear that multiplying the gyrokinetic equation by $f_0/F_M$ and integrating over the entire phase space apart from the radial direction, one can find a separate conservation equation for the intensity in the zonal modes $I_{ZM}$. In fact, applying this procedure we obtain
\be
\frac{\partial I_{ZM}}{\partial t} + \frac{\partial \Gamma_{I_{ZM}}}{\partial \psi} = C_{ZM}  \ ,
\label{intensity_ZM_conservation} 
\ee  
where $C_{ZM}$ is a source term arising from the partial integration in the non-linear term of the gyrokinetic equation given by
\be
C_{ZM} = \int d\sigma \left[ \frac{1}{F_M}\frac{\partial h_0}{\partial \psi}\sum_{n}\left(f_n\alpha_n^* \right)\right] \ ,
\label{ZM_source} 
\ee 
which can be interpreted as a correction to the source term of Eq.~\eqref{entropy_balance_terms_def}, i.e.  
this term is the flux in the gradient of the zonal perturbation, which together with the flux in the background gradient provides 
the total source.  

It is remarkable that the first term in Eq.~\eqref{flux_turbulence_split} is exactly the turbulence intensity flux connected with the zonal modes, i.e. Eq.~\eqref{intensity_ZM_conservation} shows that the zonal modes give a specific separate contribution to the turbulence intensity flux. Therefore we can now subtract Eq.~\eqref{intensity_ZM_conservation} from Eq.~\eqref{intensity_new_conservation} and obtain a conservation equation for the turbulence intensity in the perturbations $I_{P}$, i.e.
\be
\frac{\partial I_{P}}{\partial t} + \frac{\partial \Gamma_{I_{P}}}{\partial \psi} = C - C_{ZM}  \ ,
\label{intensity_P_conservation} 
\ee
this equation shows that the zonal modes enter the equation for $I_{P}$ only as a modification of the source term.

\section{CONCLUSIONS}

We have shown that starting from the conservation equation for the entropy, it is possible to write two separate conservation equations: Eq.~\eqref{intensity_ZM_conservation} describes the evolution of the turbulence intensity in the zonal modes ($I_{ZM}$) and Eq.~\eqref{intensity_P_conservation} the turbulence intensity in all the other perturbations ($I_{P}$). The turbulence flux connected to $I_{P}$, as shown in Eq.~\eqref{flux_turbulence_split}, does not receive any contribution from the zonal modes. Eq.~\eqref{intensity_P_conservation} shows that the zonal modes contribute to the conservation equation for $I_{P}$ as a correction to the source term given by the flux in the gradient of the zonal perturbation. 

This treatment gives an operative tool to actually measure the flux of turbulence in gyrokinetic numerical calculations and can therefore be used to quantitatively study the problem of the radial propagation of turbulence in tokamak plasmas.

Comparing the form of the fluxes given here with the ones derived in eq. (5) of \cite{GAR94}, it is possible to argue that $\Gamma_{I_{ZM}}$ should show the behaviour of a convective flux, since it contains only toroidal modes coupling, while $\Gamma_{I_{P}}$ should show the features of a diffusion flux, since it is given by non-linear modes coupling. 


\end{document}